\documentclass[a4paper, 11pt]{article}

\usepackage{graphicx}
\usepackage{epsf,amsmath,bbold,amsfonts,stmaryrd}
\usepackage{multirow}
\usepackage{makecell} 
\usepackage[utf8]{inputenc}
\usepackage{mathrsfs}
\usepackage{appendix}
\usepackage{amssymb}
\usepackage{float}
\usepackage{color}
\usepackage{cite}
\usepackage{hyperref}
\hypersetup{pageanchor=false}
\usepackage{indentfirst}
\usepackage{url}
\usepackage{float}
\usepackage{caption}
\usepackage[numbers,square,comma,sort&compress,merge]{natbib}

\usepackage{subfigure}

\usepackage{ulem}

\def\be{\begin{equation}}
\def\ee{\end{equation}}

\def\bea{\begin{eqnarray}}
\def\eea{\end{eqnarray}}

\def\ba{\begin{array}}
\def\ea{\end{array}}

\def\bc{\begin{center}}
\def\ec{\end{center}}

\def\bl{\begin{flushleft}}
\def\el{\end{flushleft}}

\def\br{\begin{flushright}}
\def\er{\end{flushright}}

\def\bi{\begin{itemize}}
\def\ei{\end{itemize}}

\def\bt{\begin{tabular}}
\def\et{\end{tabular}}

\newcommand{\pa}[1]{\left(#1\right)}

\numberwithin{equation}{section}

\begin{document}

\title{Testing solitonic boson star interpretations of Sagittarius A* with near-infrared flare astrometry}

\author{Xiangyu Wang$^{1}$, Zhenyu Zhang$^{2\ast}$, Hai-Qing Zhang$^{1, 3}$, Minyong Guo$^{4, 5\dagger}$, Bin Chen$^{2,6\ddagger}$
}
\date{}

\maketitle

\vspace{-10mm}

\begin{center}
{\it
	$^1$ Center for Gravitational Physics, Department of Space Science, Beihang University, Beijing
	100191, China\\\vspace{4mm}
	
	$^2$Institute of Fundamental Physics and Quantum Technology, \& School of Physical Science and Technology,
	Ningbo University, Ningbo, Zhejiang 315211, P. R. China\\\vspace{4mm}

	$^3$Peng Huanwu Collaborative Center for Research and Education, Beihang University, Beijing
	100191, China\\\vspace{4mm}

	$^4$Department of Physics, Beijing Normal University, Beijing 100875, China\\\vspace{4mm}
	
	$^5$Key Laboratory of Multiscale Spin Physics (Ministry of Education), Beijing Normal University,	Beijing 100875, China\\\vspace{4mm}

	$^6$School of Physics, \& Center for High Energy Physics, Peking University,
	No.5 Yiheyuan Rd, Beijing 100871, China\\\vspace{4mm}
}
\end{center}

\vspace{8mm}

\begin{abstract}

We use GRAVITY near-infrared (NIR) flare astrometry to test whether Sagittarius A* could be a solitonic boson star. We consider five spherically symmetric solitonic boson-star models with different effective radii, together with the Schwarzschild black hole. Treating the flares as hot spots on circular equatorial orbits, we analyze their centroid motions and images in these spacetimes and use them for parameter fitting. We perform the fitting using both $\chi^2$ analysis and Markov Chain Monte Carlo (MCMC) methods, which yield consistent results: the inferred masses of boson-star models are systematically larger than the established value of $4.3\times10^6M_\odot$. Notably, more diffusive boson stars exhibit imaging properties closer to those of a black hole, leading to mass estimates that are correspondingly closer to the established value. Overall, our results place stringent constraints on solitonic boson star interpretations of Sagittarius A*, although  do not completely rule them out.

\end{abstract}

\vfill{
	\footnotesize $\ast$ Corresponding author: zhangzhenyu@nbu.edu.cn

	\footnotesize $\dagger$ Corresponding author: minyongguo@bnu.edu.cn

	\footnotesize $\ddagger$ Corresponding author: chenbin1@nbu.edu.cn
}

\maketitle

\newpage
\baselineskip 18pt

\section{Introduction}\label{sec1}

Boson stars are among the most extensively studied horizonless compact objects and represent a prominent class of black hole (BH) mimickers. They arise as fully self-consistent, globally regular solutions to the Einstein-Klein-Gordon equations, describing gravitating configurations of a complex scalar field. By introducing self-interactions, boson stars can form ultra-compact configurations similar to black holes \cite{Cardoso:2021ehg}, and their stability and dynamical properties have been extensively studied \cite{Liebling:2012fv,DiGiovanni:2020ror}. These models have attracted considerable interest not only because they provide concrete, theoretically motivated alternatives to classical black holes, but also because they appear naturally in scenarios involving fundamental scalar fields, dark matter models, and extensions of general relativity. From an observational perspective, boson stars offer a well-posed testbed for examining how strongly gravitating but horizonless objects might imitate BH signatures, such as accretion dynamics \cite{Torres:2002td,Guzman:2005bs,Guzman:2009zz,Macedo:2013jja,Palenzuela:2017kcg,Olivares:2018abq,Olivares-Sanchez:2024dfh}, lensing structures \cite{Herdeiro:2021lwl,Rosa:2022tfv,Rosa:2022toh,Rosa:2023qcv,Zhang:2025xnl}, and gravitational waves \cite{Yunes:2016jcc,Cardoso:2016oxy,Sennett:2017etc,CalderonBustillo:2020fyi}. 

As the compact radio source at the center of the Milky Way, Sagittarius A* remains the best empirical candidate for hosting a supermassive black hole \cite{Gillessen:2008qv,Genzel:2010zy,GRAVITY:2018ofz,GRAVITY:2020gka,Do:2019txf,EventHorizonTelescope:2022wkp,Yan:2022fkr,Zhang:2024fpm}. At the same time, the possibility that it could instead be an ultracompact object has motivated substantial theoretical and observational efforts \cite{Torres:2000dw,Broderick:2009ph,Vincent:2015xta,Cardoso:2017njb,Cardoso:2017cqb,Cunha:2018gql,Zhang:2021xhp,Fromm:2021flr,EventHorizonTelescope:2022xqj,Cardoso:2019rvt}. The rapid improvement in high-resolution astronomical instruments over the past decade has made it increasingly feasible to confront such alternatives with data. A major breakthrough came from the GRAVITY Collaboration, whose NIR observations have dramatically improved our understanding of Sgr A* flares. 

In 2018, the GRAVITY Collaboration reported continuous astrometric measurements of three NIR flare events from Sgr A* \cite{Abuter:2018uum}. More recently, the 2023 GRAVITY data set expanded the sample to four astrometric and six polarimetric flare events \cite{GRAVITY:2023avo}. All observed flares display consistent clockwise, loop-like motion with a characteristic angular scale of about 150 $\mu$as and durations of tens of minutes. Simultaneously, the polarization angle was observed to rotate coherently with a similar period. These measurements provide time-resolved dynamical constraints on the spacetime geometry in the vicinity of Sgr A*, making them highly suitable for testing BH alternatives such as boson stars \cite{Rosa:2025dzq,Aimar:2025uia}. The hotspot model provides a simple and effective theoretical explanation for these flare phenomena \cite{Broderick:2005jj,Vincent:2023sbw,Huang:2024wpj,Chen:2024ilc}, where the flares originate from high-temperature plasmoids near the central object \cite{genzel2010galactic,Ripperda:2020bpz,Ripperda:2021zpn,Aimar:2023kzj,Porth:2020txf,Najafi-Ziyazi:2023oil,Chen:2024ggq,Zhao:2025ouq,Dexter:2020cuv,Jiang:2024gtk}. Relatively localized plasmoids can be treated as point sources during the imaging process, and can therefore be conveniently used to fit the GRAVITY data \cite{GRAVITY:2020lpa,Ball:2020jup,Xie:2025skg,Matsumoto:2020wul,Antonopoulou:2024qco,Yfantis:2024eab}.

In this study, we use GRAVITY NIR flare astrometry data to test whether Sgr A* could be a solitonic boson star. We consider five spherically symmetric boson-star models with different compactness, and obtain their metrics numerically. By interpreting flares as hotspots on circular geodesics, we employ general relativistic radiation transfer (GRRT) to generate dynamical images for each model. From these images, we compute the time evolution of the intensity centroid and compare it with observations. We use the grid-$\chi^2$ method to constrain the orbital parameters corresponding to the combined data. For comparison, we also use MCMC combined with Bayesian inference to constrain the global parameters shared by three individual events. The results from both methods show the same trend: the boson stars with a more diffuse distribution of mass have masses closer to the well-established mass of Sgr A*, meaning that their confidence intervals are more likely to contain this reference value. These findings impose meaningful constraints on boson star interpretations, while still leaving some viable parameter space.

The remaining part of this paper is organized as follows. In Sec.~\ref{sec2}, we review the numerical solutions of solitonic boson stars, outlining the configurations used in our calculations. In Sec.~\ref{sec3}, we introduce the hotspot-emission model and briefly review the GRRT framework. In Sec.~\ref{sec4}, we outline the fitting strategy, including both the $\chi^2$-analysis for combined data and the MCMC procedure for global parameter estimation. In Sec.~\ref{sec5}, we present the fitting results and discuss their implications for the nature of Sgr A*. Finally, we conclude our study and provide an outlook for future research in Sec.~\ref{sec6}.

\section{Boson star solutions}\label{sec2}
In this section, we display the approach for solving boson stars and the specific configurations employed in our study. The action governing the coupled scalar-gravitational fields can be expressed as
\begin{equation}
\mathcal{S} = \int d^4x \sqrt{-g} \left[ \frac{R}{2\kappa} - \frac{1}{2} \nabla_a \Phi^* \nabla^a\Phi - V(|\Phi|^2) \right],
\end{equation}
where $g$ denotes the determinant of the spacetime metric $g_{\mu\nu}$. $R$ is the Ricci scalar and $\kappa=8\pi G$ is the coupling constant. We adopt natural units with $G=c=1$ here, so that
$\kappa=8\pi$ in the later calculation. The complex scalar field is denoted as $\Phi$ with $\Phi^*$ indicating its conjugate field. Function $V(\Phi)$ represents the potential of this scalar field. In this study, we consider the scalar potential of the form 
\begin{equation}
V = \mu^2 |\Phi|^2 \left(1 + \frac{|\Phi|^2}{\alpha^2}\right)^2,
\end{equation}
where $\mu$ represents the boson mass and $\alpha$ is a constant parameter.

The equations of motion can be derived by applying the variational principle to the metric and fields,
\bea
\label{gphieoms}
\begin{aligned}
R_{ab}-\frac{1}{2}g_{ab}R=\kappa T_{ab},\,\\
\nabla_a\nabla^a\Phi=\Phi\frac{dV}{d|\Phi|^2},\,
\end{aligned}
\eea
where $T_{ab}$ is the energy-momentum tensor, defined as 
\begin{equation}
T_{ab}\equiv\nabla_a\Phi^*\nabla_b\Phi+\nabla_b\Phi^*\nabla_a\Phi-g_{ab}(\nabla_c\Phi^*\nabla^c\Phi+V).
\end{equation}
In this paper, we consider a spherically symmetric boson star spacetime, whose line element can be written as 
\be
\label{dsansatz}
ds^2 = -A dt^2 + B^{-1} dr^2 + r^2 (d\theta^2+\sin^2\theta d\phi^2),\, 
\ee
and scalar field ansatz is given by
\be
\label{phiansatz}
\Phi = \psi(r) e^{-i\omega t}.
\ee
Substituting Eq. \eqref{dsansatz} and Eq. \eqref{phiansatz} into the equations of motion Eq. \eqref{gphieoms} yields
\bea
\label{BSeom}
\begin{aligned}
A'&=\frac{A(1-B)}{Br}+\kappa r\left(\frac{\omega^2\psi^2}{B}+A(\psi')^2-\frac{A}{B}V\right),\\
B'&=\frac{1-B}{r}-\kappa r\left(\frac{\omega^2\psi^2}{A}+B(\psi')^2+V\right),\\
\psi''&+\left(\frac{2}{r}+\frac{1}{2}\left(\frac{A'}{A}+\frac{B}{B'}\right)\right)\psi'+\frac{1}{B}\left(\frac{\omega^2}{A^2}-\frac{dV}{d\psi}\right)\psi=0.
\end{aligned}
\eea
\begin{figure}[h!]
	\centering
	\includegraphics[width=\textwidth]{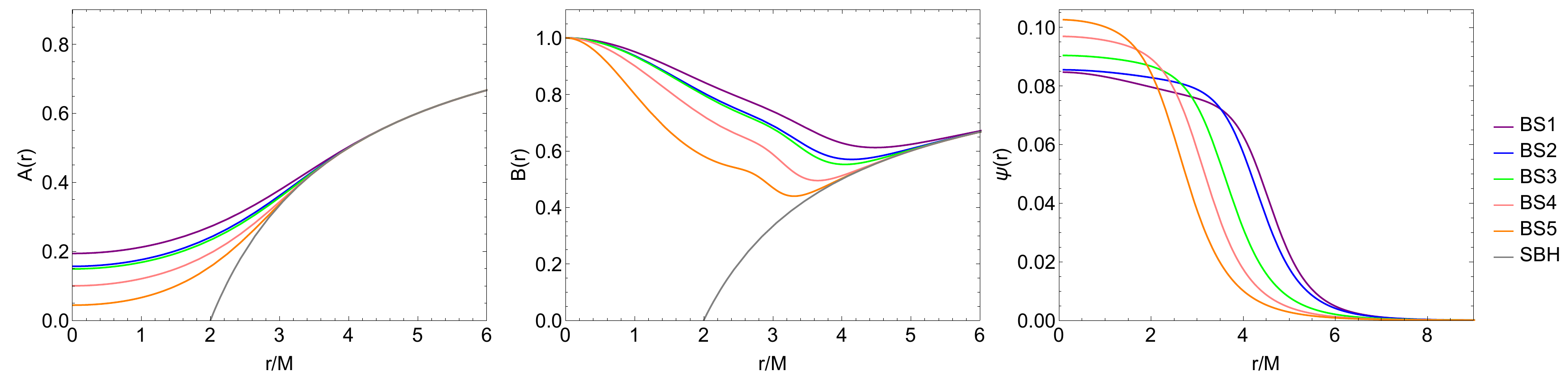}
	\caption{The metric and scalar field solutions for the BS1 (purple), BS2 (blue), BS3 (green), BS4(red) and BS5(orange). The metric components as functions of $r$ are shown in the left panel and the middle panel, while the gray lines mark the reference function $1-2M/r$. The right panel shows the spatial distribution of the  scalar field $\psi(r)$.}
	\label{so}
\end{figure}

To obtain the boson star solution, we also need to specify appropriate boundary conditions. At the origin, the boundary conditions are
\bea
\begin{aligned}
	A&(r\rightarrow0)=A_0,\\
	B&(r\rightarrow0)=1,\\
	\psi&(r\rightarrow0)=\psi_0.
	\label{bound}
\end{aligned}
\eea
The boundary conditions at infinity are given by the Schwarzschild solution,
\bea
\begin{aligned}
A&(r\rightarrow\infty)=1-\frac{2M}{r},\\
B&(r\rightarrow\infty)=1-\frac{2M}{r},\\
\psi&(r\rightarrow\infty)=0,
\end{aligned}
\eea
where $M$ represent the Arnowitt-Deser-Misner (ADM) mass of the boson star.

We can then employ the shooting method to solve Eq.~\eqref{BSeom}. Since the differential equations are independent of time $t$, the temporal dimension can be freely rescaled. In the actual solution process, we first set $A_0$ to a fixed value to solve the equations, and then reparameterize $A_0$ to satisfy the boundary conditions at infinity. Therefore, in practice, the only free parameters during the solution process are $\alpha$ and $\psi_0$. A detailed discussion of this method can be found in \cite{Ma:2025iti}. 

\begin{table}[!htbp]
	\centering
	\renewcommand{\arraystretch}{1.3}
	\begin{tabular}{l|c|c|c|c|c|c}
		\hline\hline
		Model & $\alpha$ &$\psi_0$ & $\mu M$ & $\mu R$ & $\omega/\mu$ & $R/M$ \\
	    \hline 
		BS1 & 0.090 & 0.0103 & 0.742 & 4.20 & 0.438 & 5.66 \\ 
		\hline 
		BS2 & 0.085 & 0.0969 & 0.916 & 4.54 & 0.385 & 4.95 \\ 
		\hline
		BS3 & 0.080 & 0.0907 & 1.063 & 4.96 & 0.353 & 4.66  \\ 
		\hline
		BS4 & 0.075 & 0.0854 & 1.348 & 5.56 & 0.303 & 4.12 \\ 
		\hline 
		BS5 & 0.070 & 0.0847 & 1.530 & 5.55 & 0.273 & 3.62 \\ 
	
		\hline\hline
	\end{tabular}
	\caption{Parameters for different boson star model.}
	\label{para}
\end{table}

In this paper, we consider sufficiently compact boson stars that can produce higher-order photon ring structures similar to those in black hole imaging. We examine five different cases: $\alpha = 0.090, \psi_0 = 0.0103$(denote as BS1), $\alpha = 0.085, \psi_0 = 0.0969$ (denoted as BS2),  $\alpha = 0.080, \psi_0 = 0.0907$ (denoted as BS3), $\alpha = 0.075, \psi_0 = 0.0854$ (denoted as BS4), and $\alpha = 0.070, \psi_0 = 0.0847$ (denoted as BS5), all of which satisfy the condition. The corresponding $A$, $B$, and $\psi$ are shown in Fig. \ref{so}, with the relevant parameters summarized in Tab. \ref{para}. As shown in the figure, when $r$ reaches about $5M$, all boson-star metric components converge to the Schwarzschild case, as expected. Meanwhile, the boson-star configurations become progressively more compact from BS1 to BS5. To characterize this behavior, it is useful to define the effective radius via
\begin{equation}
B(r)=1-\frac{2m(r)}{r},\quad m(R) = 0.98M.
\end{equation}
For regions with $r > R$, the spacetime can essentially be treated as Schwarzschild metric. We list the corresponding effective radii of boson stars in the last column of Tab.~\ref{para}.

\section{Emission model and imaging process}\label{sec3}
In this section, we will introduce the hot-spot emission model and the radiation transfer method employed in our study. We consider hot spots undergoing Keplerian motion, that is, their trajectories correspond to timelike circular geodesics in spacetime. For spherically symmetric spacetimes, the four-velocity $u^{\mu}$ of circular orbits on the equatorial plane is given by
\begin{equation}
u^{\mu} = \zeta(1, 0, 0, \Omega)\,, \quad \zeta = \sqrt{\frac{-1}{-A(r) + r^2\Omega^2}}\,,
\end{equation}
where $\Omega \equiv d\phi/dt$ is the angular velocity. For Keplerian motion, it takes the form
\begin{equation}
\Omega_{\pm} =\pm\sqrt{\frac{-\partial_r A(r)}{2r}}.
\end{equation}
The $\pm$ solutions correspond to prograde and retrograde orbits respectively. Then the four-velocity of hot spots can be obtained by substituting the corresponding metric components into these equations. 

We then proceed to perform imaging using the general-relativistic ray-tracing and radiative transfer technique. For details of the GRRT scheme we adopt, see \cite{Zhang:2024lsf,Huang:2024bar}. For simplicity, we model the hotspot as an optically thin sphere with negligible absorption. In this case, the radiative transfer equation along a null geodesic can be written as
\bea
\frac{d}{d\lambda}\left(\frac{I_\nu}{\nu^3}\right)=\frac{j_\nu}{\nu^2},
\eea
where $I_\nu$ is the intensity, $j_\nu$ is the emissivity, $\nu$ is the photon frequency measured in the local rest frame, and $\lambda$ is the affine parameter along the photon trajectory. We further assume a flat emission spectrum, namely $j_\nu = J$ with $J$ a frequency‑independent emissivity coefficient. Then the frequency dependence factors out, and after integrating the transfer equation from the source to the observer we obtain the observed intensity
\bea
I_o = \nu_o \int g^2 \, J \, d\lambda,
\eea
where $g \equiv \nu_o/\nu_e$ is the redshift factor relating the observed frequency $\nu_o$ to the emitted frequency $\nu_e$. Similar to \cite{GRAVITY:2020lpa,Zhang:2025vyx}, we assume that the emissivity of the hotspot exponentially decays with distance from its center, satisfying
\bea
J\propto e^{\frac{-\boldsymbol{x}^2}{2s^2}},
\eea
where $|\boldsymbol{x}|$ is the spatial distance from the emitting point to the hotspot center, and $s$ is a physical quantity characterizing the size of the hotspot itself. For an emitting point with coordinates $\pa{t_{p},r_{p},\theta_{p},\phi_{p}}$, we have
\bea
\begin{aligned}
\boldsymbol{x}^2=\pa{r_{p} \sin\theta_{p}\cos\phi_{p} - r_{\text{hs}}\cos\Omega_{\text{hs}}t}^2+
\pa{r_{p} \sin\theta_{p}\sin\phi_{p} - r_{\text{hs}}\sin\Omega_{\text{hs}}t}^2+r_{p}^2 \cos^2\theta_{p},
\end{aligned}
\eea
where $r_\text{hs}$ is the orbital radius of the hotspot and $\Omega_\text{hs}$ is the corresponding angular velocity. Since $\boldsymbol{x}^2$ depends on both space and time, the emissivity $J$ describes a dynamically evolving source system. In the following numerical imaging scheme, we set $s=0.3M$.

\begin{figure}[h!]
	\centering
	\includegraphics[width=0.4\textwidth]{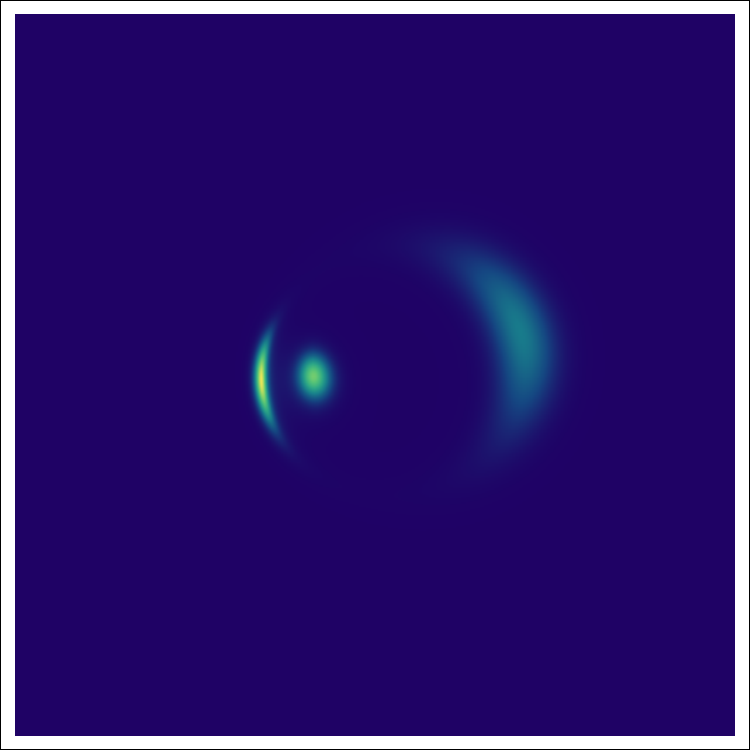}
	\caption{A snapshot of a hot spot orbiting a boson star. The hot spot radius $r_{\text{hs}} = 7M$, and the inclination $\theta_o = 80^\circ$. }
	\label{ex}
\end{figure}

As an example, Fig. \ref{ex} shows a single frame of a hot-spot movie orbiting BS1. The orbital radius of the hot spot is set to $r_{\text{hs}} = 7M$, and the observer is located at an inclination angle of $\theta_o = 80^\circ$. From the image, three distinct features can be clearly seen: the primary image having the largest area, the secondary image resembling an arc, and the small bright spot closest to the center of the image. The central bright spot is formed by light rays passing through the boson star and is the main difference between boson star images and black hole images, which is consistent with previous studies on boson star imaging \cite{Rosa:2022toh,Rosa:2023qcv}.

After obtaining a series of intensity distribution snapshots, we may calculate the variation of the centroid position in order to compare it with the flare astrometry observed by GRAVITY. The flux of a single pixel is given by 
\begin{equation}
F(i,j) = I_o(i,j) S_0 \cos \Psi_{\text{in}}(i,j),
\end{equation}
where $I_o(i,j)$ is the intensity of the pixel $(i,j)$, $S_0$ represents the pixel size and 
$\Psi_{\text{in}}$ denotes the incident angle of the light ray relative to the imaging plane. 
For the camera model we used, the incident angle $\Psi_{\text{in}}$ takes the form \cite{Huang:2024wpj}
\begin{equation}
\Psi_{\text{in}}(i,j) = 2 \arctan \left(\frac{2}{N} \tan \left( \frac{\alpha_{\text{fov}}}{2} \right) \sqrt{ \left( i - \frac{N+1}{2} \right)^2 + \left( j - \frac{N+1}{2} \right)^2 } \right),
\end{equation}
where $N$ is the total number of pixels in each column or row, and $ \alpha_{\text{fov}} $ is the field of view angle. Then, the centroid of emission $\boldsymbol{X}_c $ can be obtained as the weighted average of the position with the flux serving as the weight, namely,
\begin{equation}
\boldsymbol{X}_c = \frac{\sum\limits_{i,j}\boldsymbol{X}(i,j) F(i,j)}{\sum\limits_{i,j} F(i,j)},
\end{equation}
where $\boldsymbol{X}_c=\left(X,\,Y\right)$ denotes the coordinates in the observational plane.

\section{Fitting process}\label{sec4}
In this section, we introduce the methods used for fitting. Since we consider parameter fitting under several fixed boson-star spacetime backgrounds, there are five parameters to be fitted, namely,
\begin{equation}
    \boldsymbol{\Theta}=(r_\text{hs},\, \theta_o,\,\text{PA},\, \phi_0\,,M),
\end{equation}
where $r_\text{hs}$, $\theta_o$ and $M$ denote, respectively, the hotspot obital radius, inclination and mass of the boson star, as mentioned earlier. In addition, PA represents the position angle, and $\phi_0$ represents the initial azimuthal coordinate of the hot spot. 
Among these parameters, $r_\text{hs}$ and $\phi_0$ characterize the orbital properties of the hotspot and thus vary in different flare events. In contrast, $\theta_o$, PA, and $M$ describe the observer’s orientation and the spacetime itself, and therefore should share identical values across all events. Consequently, we can either analyze each flare event individually to constrain all five parameters, or perform a joint analysis of multiple flare events to constrain the three global parameters. Different fitting approaches are adopted for these two cases.

\subsection{Grid $\chi^2$}

For a single flare event, our approach involves discretizing the parameter space into a grid and computing the $\chi^2$ at each grid point to identify the best-fitting parameter set. For each point in parameter space, the $\chi^2$ is given by:
\begin{equation}
\chi^2(\boldsymbol{\Theta}) \equiv \sum_i \left[ \frac{\left( X_c^i(\boldsymbol{\Theta}) - X_o^i \right)^2 }{ \sigma_{X^i}^2 }+\frac{\left( Y_c^i(\boldsymbol{\Theta}) - Y_o^i \right)^2 }{ \sigma_{Y^i}^2 } \right],
\end{equation}
where the subscript $c$ denotes the centroid position from the model, i.e., calculated by the GRRT method introduced in Sec. \ref{sec3}. The subscript $o$ denotes the observational data and the superscript $i$ corresponds to the astrometric time $t_i$. $\sigma_{X^i}$ and $\sigma_{Y^i}$ denote the standard deviations of $X^i$ and $Y^i$, which are given by the observational uncertainties. Additionally, $N_d$ denotes the number of data points and $N_f=5$ denotes the number of parameters. The parameter set \( \boldsymbol{\Theta} \) that minimizes $ \chi^2 $ is regarded as the optimal model fit.

To summarize, we present the ranges of the parameters along with the corresponding number of grid points in the Tab.~\ref{range}. We only consider cases with $\theta_o > 90^\circ$. This is because the hot spots orbit in the direction of increasing $\phi$ in our setup, while the flares observed by GRAVITY all exhibit clockwise motion. This condition is satisfied when the observer is located in the lower hemisphere. Owing to the spherical symmetry of the metric, this configuration is entirely equivalent to the case of an observer with $\theta_o < 90^\circ$ observing a hot spot orbiting in the direction of decreasing $\phi$. 

\begin{table}[h]
\centering
\begin{tabular}{c|c|c|c}
\hline\hline
Parameter & Meaning & Range &  grid points \\
\hline
$r_{\text{hs}}$ & orbital radius  & $[4M, 12M]$ & 17 \\
$\theta_o$ & inclination & $[90^\circ,\,180^\circ]$ & 19 \\
$\phi_0$ & initial azimuth & $[0, \,360^\circ]$ & 36 \\
PA & position angle & $[0, \,180^\circ]$ & 36\\
$M$ & ADM mass & $[2\times10^6 M_\odot, \,12\times10^6 M_\odot]$ & 51 \\
\hline
\end{tabular}
\caption{The parameters with their ranges and the number of grid points.}
\label{range}
\end{table}
\par

For a fixed metric, we need to generate black hole movies for different values of $r_\text{hs}$ and $\theta_o$. Each movie must cover one full orbital period of the hot spot, and we divide each period into 20 frames. Since we will later consider six different metrics i.e., five boson stars and the Schwarzschild black hole), and given the grid numbers described in the Tab.~\ref{range}, we need to use GRRT to generate $17\times19\times20\times6=38760$ images in total. After that, for different values of $\phi_0$, PA, and $M$, we can obtain the theoretical dynamic images of the hot spot by applying time translation, rotations, and overall rescaling.

This grid-based evaluation of the likelihood function via the metric allows us to rigorously determine the optimal model parameters within a Bayesian framework. From a Bayesian perspective, the parameters $\boldsymbol{\Theta}$ can be viewed as a set of random variables which evidently follow a posterior probability distribution $P(\boldsymbol{\Theta})$. The posterior probability $P(\boldsymbol{\Theta})$ is proportional to the product of the likelihood function and the prior probability distribution function. In our framework, uniform prior distributions are assumed for all parameters, and the likelihood function is typically assumed to have a Gaussian form. Therefore, its relationship with the $\chi^2$ is expressed as:
\begin{equation}
\mathcal{L}_{\text{Likelihood}}(\boldsymbol{\Theta})=\exp\left[-\frac{1}{2}\chi^2(\boldsymbol{\Theta})\right]
\label{singlelike}
\end{equation}
Consequently, by using the $\chi^2$ value that we obtained via the grid-based method introduced in the above section, we are able to calculate the posterior probability. The marginalized distributions of the parameters are obtained through grid-based numerical integration. To characterize the distribution of the posterior probability, we compute the Highest Density Interval (HDI) using the \textit{ArviZ} package in Python \cite{arviz_2019}. The HDI serves as a commonly used measure to summarize the dispersion of a posterior distribution, defined as the shortest interval encompassing a given probability density.

\subsection{Markov Chain Monte Carlo}

For the case of constraining the global parameters using multiple events, the grid-based method introduced above is not suitable. This is because, in addition to the three global parameters to be fitted, we also need to consider the individual internal parameters $r_{\text{hs}}$ and $\phi_0$ for each event in the calculation. This means that the actual dimension of the parameter space increases by $2n$ with $n$ the number of flare events included, which can easily lead to the curse of dimensionality and an explosion in computational cost. Therefore, we need to employ the Markov Chain Monte Carlo (MCMC) method to fit the parameters. 

In this process, the total likelihood is defined as the product of the likelihoods of all individual events, which is equivalent to summing their contributions in the exponent:
\begin{equation}
\mathcal{L}_{\text{tot}}
= \prod_k \mathcal{L}_k(\boldsymbol{\Theta}_k)
= \exp\left[-\frac{1}{2}\sum_k \chi_k^2(\boldsymbol{\Theta}_k)\right],
\end{equation}
where $\mathcal{L}_k(\boldsymbol{\Theta}_k)$ is given by Eq.~\eqref{singlelike} for the data of the $k$-th event. Note that the different parameter sets $\boldsymbol{\Theta}_k$ include the same three global parameters. 

Besides, we usually require $\mathcal{L}_\text{tot}$ to be a continuous function of the parameter set in MCMC, so interpolating the previously grid-computed results is necessary. Specifically, we compute the gridded centroid positions from the GRRT results, and then interpolate to obtain continuous functions $X_c(r_\text{hs},\theta_o,t_\text{obs})$ and $Y_c(r_\text{hs},\theta_o,t_\text{obs})$. The remaining parameters, PA, $\phi_0$ amd $M$, can then be incorporated through continuous transformations. After that, we can carry out the analysis using \textit{emcee} package in Python \cite{foreman2013emcee}.

\section{Results}\label{sec5}

Based on the GRRT and parameter constraint methods introduced in the previous two sections, we present our results in this section. We mainly focus on the astrometry of three flare events observed on May 27, July 22, and July 28, 2018, whose observational data were published in \cite{Abuter:2018uum, GRAVITY:2023avo}. Additionally, GRAVITY collaboration assumes that different events can be averaged to suppress the random features that are independent across events \cite{GRAVITY:2023avo}. In other words, the dynamical parameters of the hot-spot orbital motion, i.e. $r_\text{hs}$ and $\phi_0$, can be treated as mean parameters of the system. Based on this assumption, they propose a fitting method using the combined data obtained by averaging multiple events. In this paper, we also analyze this combined data.

\begin{figure}[!htbp]
	\centering
	\includegraphics[width=0.95\textwidth]{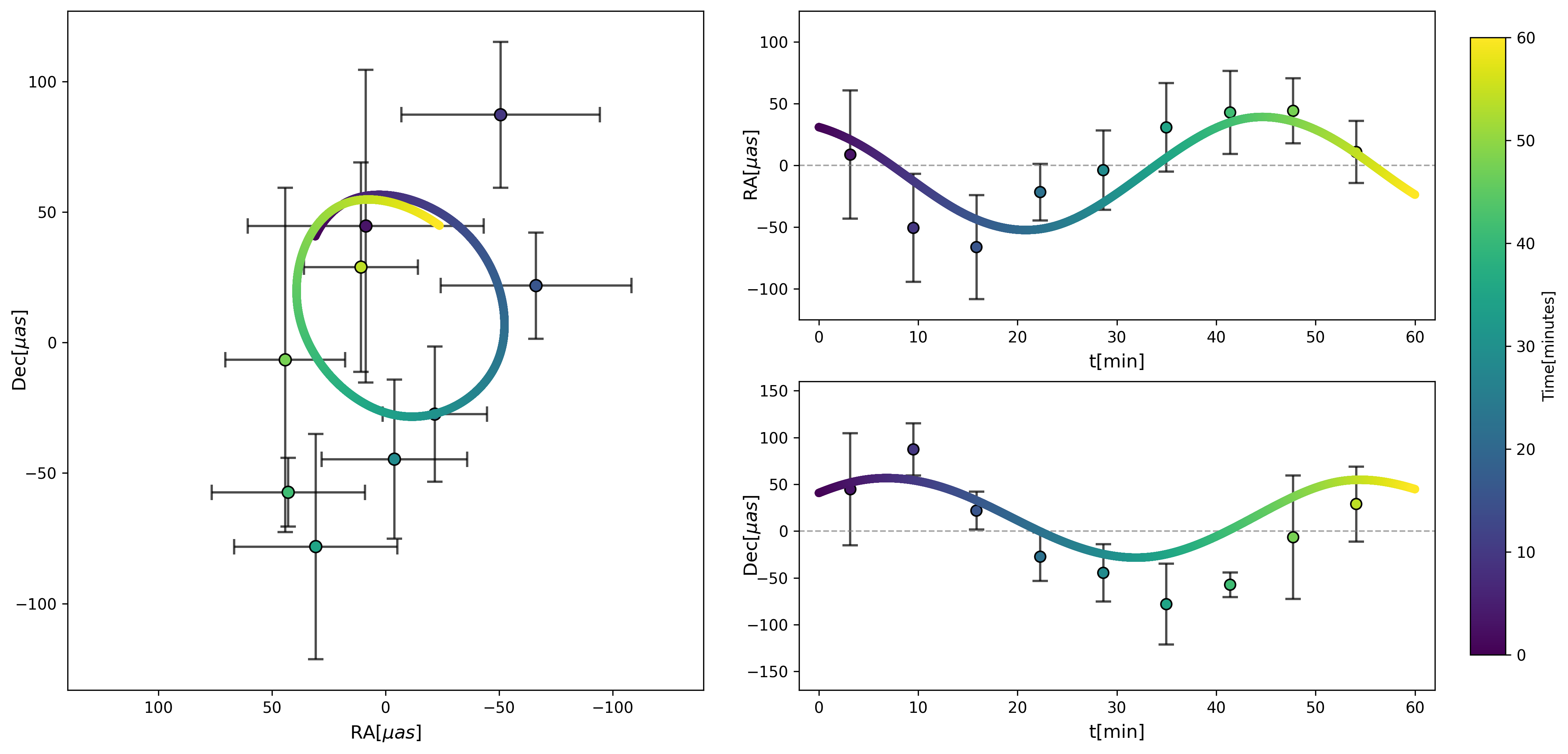}
	\caption{The best fit of BS5 model for combined data. Left: the centroid motion. Right: the Right Ascension and Declination on the sky as a function of time.}
	\label{chires}
\end{figure}

\begin{figure}[!htbp]
	\centering
	\includegraphics[width=0.5\textwidth]{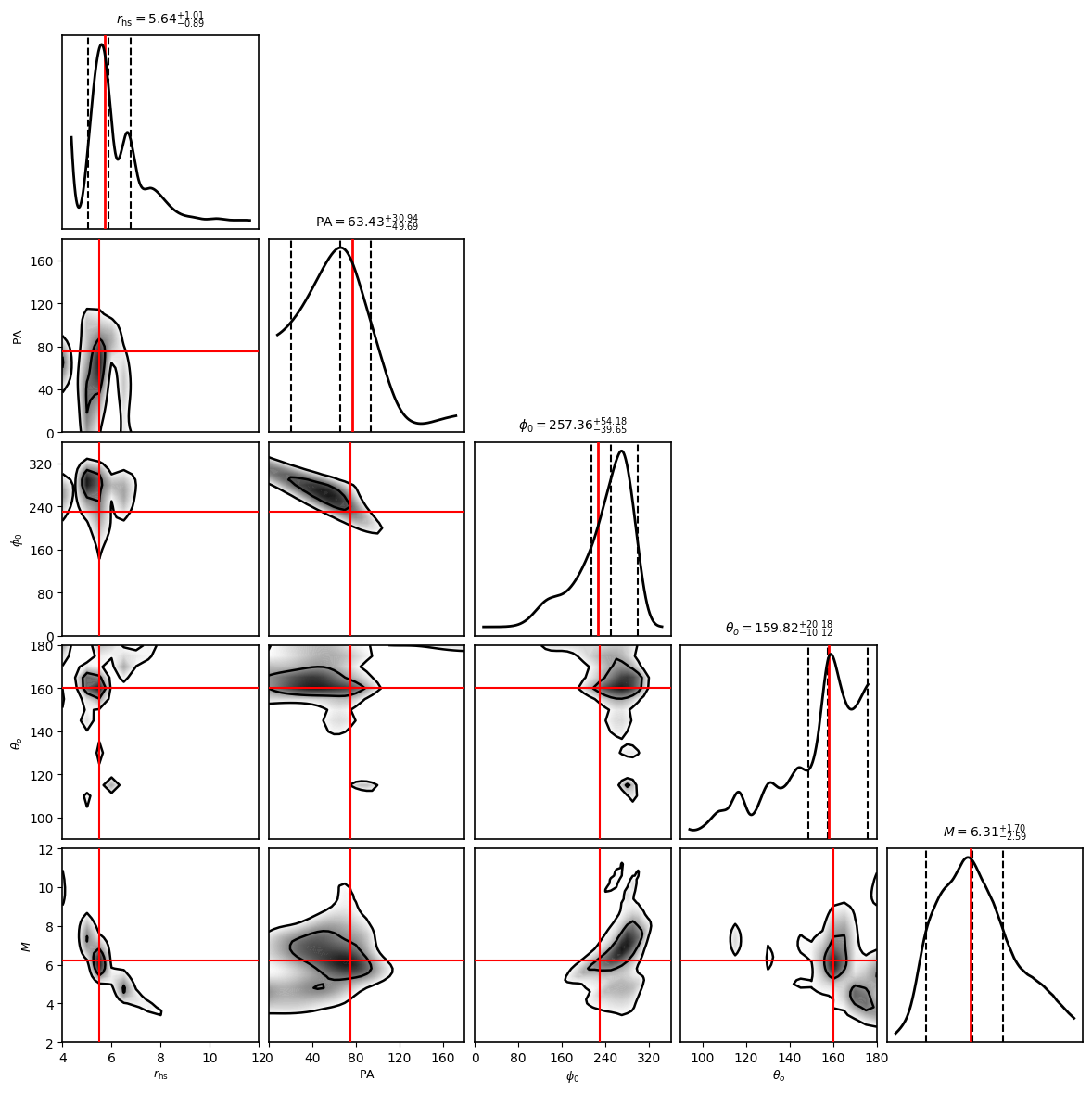}
	\caption{Posterior probability distribution functions for BS5 parameters of the combined astrometric data. The dashed lines in the diagonal plots represent the median values and the $68.3\%$ HDI of each parameter. The contours in the two-dimensional joint probability distribution plots correspond to the $39.4\%$ and $86.4\%$ HDI. The red lines denote the best-fit values. }
	\label{bysBS2}
\end{figure}

\subsection{Combined data}

We first provide the fitting results of the five parameters for the combined data.  As an example, we present the best-fit result of the BS5 model based on the minimum-$\chi^2$ fitting in Fig. \ref{chires}. The left panel displays the best-fit astrometry, with color indicating the time, while the right panel demonstrates the temporal evolution of the centroid position in the RA and Dec directions. The discrete points and line segments represent the observational data and their uncertainties, whereas the continuous curves denote the best-fit centroid motion, corresponding to
$r_\text{hs}=5.5M$, $\theta_o=160^\circ$, $\phi_0=230^\circ$, $\text{PA}=160^\circ$, $M=6.2\times10^6M_\odot$. It can be seen that the data points are almost evenly distributed on both sides of the best-fit curve. 

The Bayesian analysis of this fitting is shown in Fig.~\ref{bysBS2}. The panels along the diagonal provide the marginal probability distributions of the individual parameters, with the central dashed line marking the median value, namely the parameter value at which the cumulative distribution function of the marginal posterior reaches $50\%$. The two flanking dashed lines indicating the $68.3\%$ HDI, meaning that each parameter has a $68.3\%$ probability of falling within this interval. The panels in the lower triangle present the joint probability distributions for pairs of parameters, where the $39.4\%$ and $86.4\%$ HDIs are outlined by contours. The parameter values corresponding to the minimum $\chi^2$ are marked with red lines in each plot. We can see that the parameters $\phi_0$ and $M$ are relatively well constrained, each exhibiting a clear single peak. In contrast, the data provide weaker constraints on the remaining parameters, which display either multi-peaked structures or overly broad maxima.

\begin{table}[!htbp]
    \centering
    \renewcommand{\arraystretch}{1.3}
    \begin{tabular}{l|l|c|c|c|c|c|c}
        \hline\hline
        \multicolumn{2}{c|}{Model} & $r_\text{hs}[M]$ & $\theta_o[^\circ]$ & $\phi_0[^\circ]$ & PA$[^\circ]$ & $M[10^6M_\odot]$ & $\chi_{\text{min}}^2$ \\
        \hline
        \multirow{2}{*}{BS1} 
        & Fit & 7.5 & 130 & 135 & 135 & 4.8 & 1.21 \\ \cline{2-8}
        & HDI & $7.45^{+0.63}_{-1.32}$ & $124.86^{+15.48}_{-11.10}$ & $120.24^{+55.40}_{-47.39}$ & $136.04^{+39.81}_{-76.75}$ & $5.45^{+2.63}_{-2.92}$ & --\\\hline
        \multirow{2}{*}{BS2} 
        & Fit & 6.0 & 125 & 310 & 5 & 5.0 & 1.28 \\ \cline{2-8}
        & HDI & $6.07^{+0.52}_{-2.34}$ & $123.43^{+29.53}_{-17.28}$ & $338.55^{+38.32}_{-82.50}$ & $78.67^{+49.32}_{-74.34}$ & $5.78^{+2.61}_{-3.52}$ & --\\
        \hline
        \multirow{2}{*}{BS3} 
        & Fit & 6.5 & 150 & 330 & 5 & 5.0 & 1.32 \\ \cline{2-8}
        & HDI & $6.41^{+0.89}_{-2.41}$ & $153.10^{+24.47}_{-10.26}$ & $255.05^{+76.47}_{-42.27}$ & $67.22^{+20.38}_{-62.50}$ & $5.70^{+4.04}_{-2.69}$ & --\\ 
        \hline
        \multirow{2}{*}{BS4} 
        & Fit & 4.5 & 175 & 270 & 25 & 6.6 & 1.12 \\ \cline{2-8}
        & HDI & $5.28^{+0.51}_{-1.01}$ & $136.34^{+37.07}_{-21.09}$& $314.91^{+45.09}_{-26.81}$ & $167.00^{+13.00}_{-9.80}$  & $6.55^{+1.53}_{-1.55}$ & --\\ 
        \hline
        \multirow{2}{*}{BS5} 
        & Fit & 5.5 & 160 & 230 & 160 & 6.2 & 1.18 \\ \cline{2-8}
        & HDI & $5.64^{+1.01}_{-0.88}$ & $159.82^{+20.18}_{-10.12}$ & $257.36^{+54.18}_{-39.65}$ & $63.43^{+30.94}_{-49.69}$ & $6.31^{+1.70}_{-2.59}$ & --\\ 
        \hline
        \multirow{2}{*}{SBH} 
        & Fit & 10.5 & 110 & 100 & 105 & 4.8 & 1.0 \\ \cline{2-8}
        & HDI & $8.44^{+1.52}_{-2.19}$ & $121.19^{+20.67}_{-24.73}$ & $164.92^{+98.34}_{-136.55}$ & $112.84^{+25.78}_{-23.24}$ & $4.97^{+1.39}_{-2.44}$ & -- \\ 
        \hline\hline
    \end{tabular}
    \caption{Comparison between best-fit parameter values and posterior median values (68.3\% HDI) of comebined data for different compact object models.}
    \label{tab:combineddata}
\end{table}

We then repeat the above analysis for the other boson star models and summarize the results  in Tab.~\ref{tab:combineddata}. For comparison, we also perform the same analysis for the Schwarzschild black hole (denoted as SBH). Similar to the treatment in \cite{GRAVITY:2020gka}, since all models exhibit relatively large $\chi^2$ values, we normalize the $\chi^2$ of the Schwarzschild case to 1. We find that the minimum $\chi^2$ values of the other models are all larger than that of the Schwarzschild black hole, which may indicate that the Schwarzschild black hole provides a better fit to the astrometry data than the boson star models. In addition, compared to the black hole case, all boson star models yield larger ADM masses and smaller hotspot orbital radii. The only exception is the most diffuse model, BS1, for which the mass corresponding to the minimum $\chi^2$ is identical to that of the Schwarzschild black hole, although the median value of its mass HDI remains higher than that of the Schwarzschild case.

\begin{figure}[!htbp]
	\centering
	\includegraphics[width=0.38\textwidth]{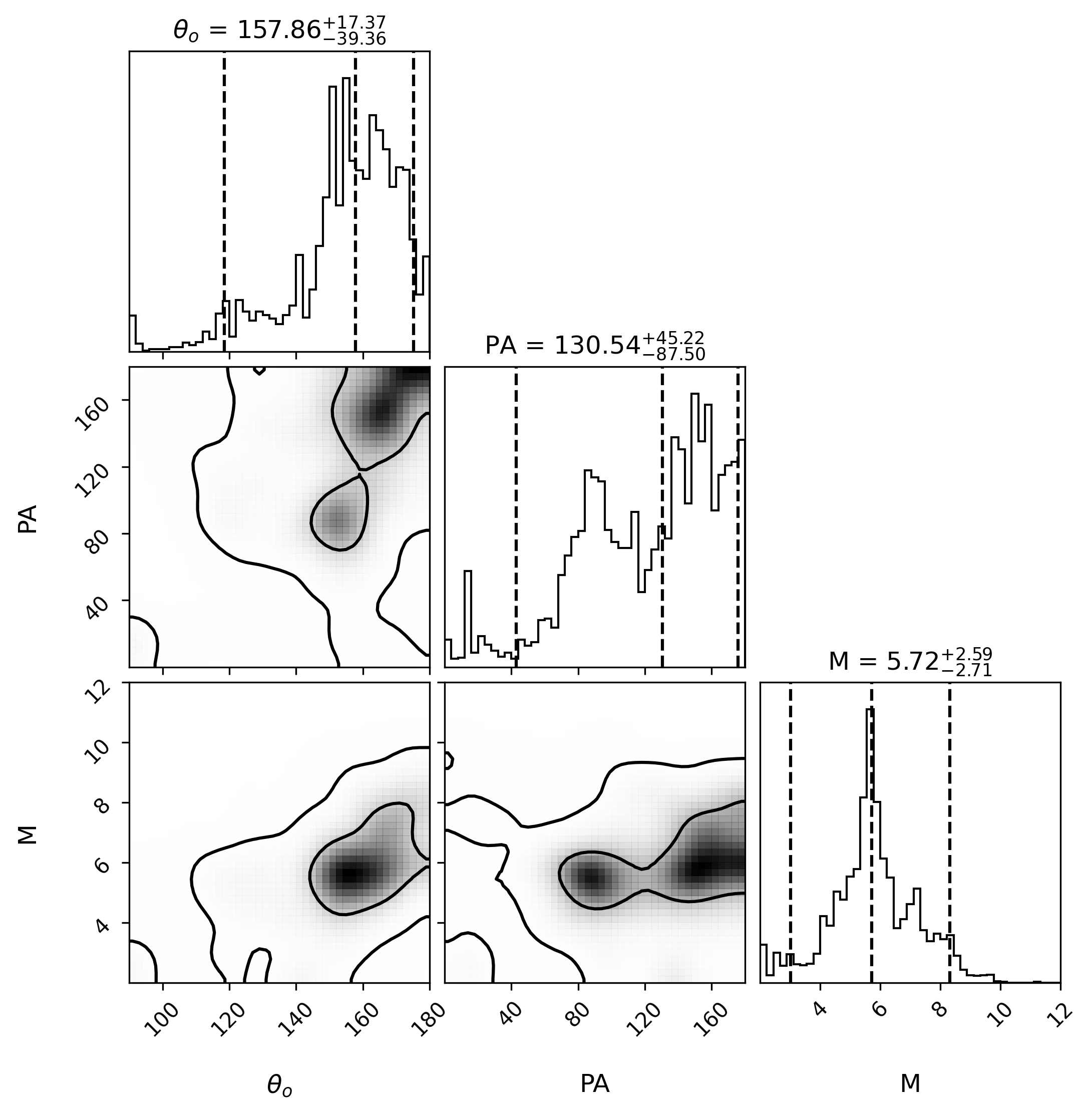}
	\includegraphics[width=0.38\textwidth]{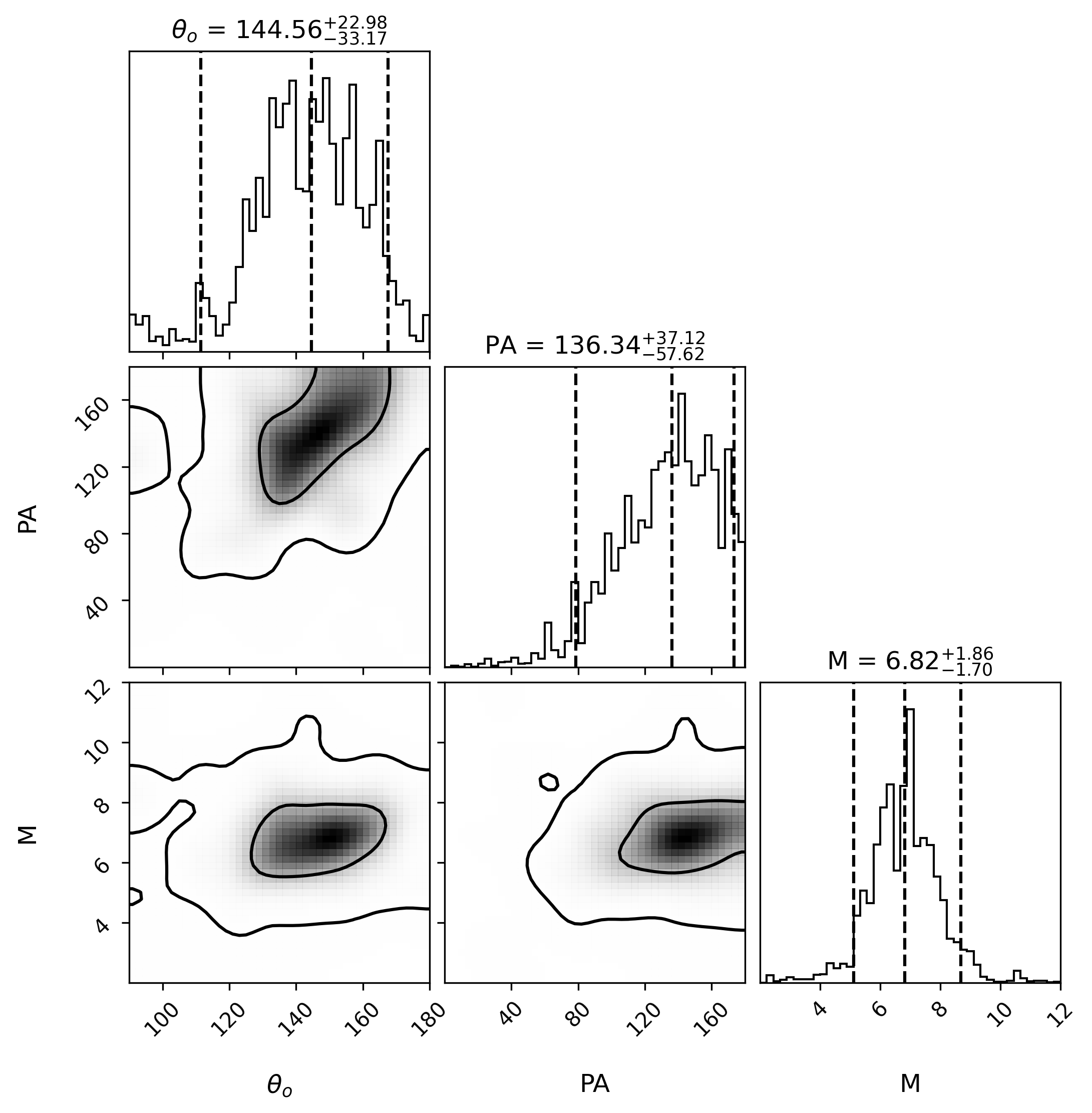}\\
	\includegraphics[width=0.38\textwidth]{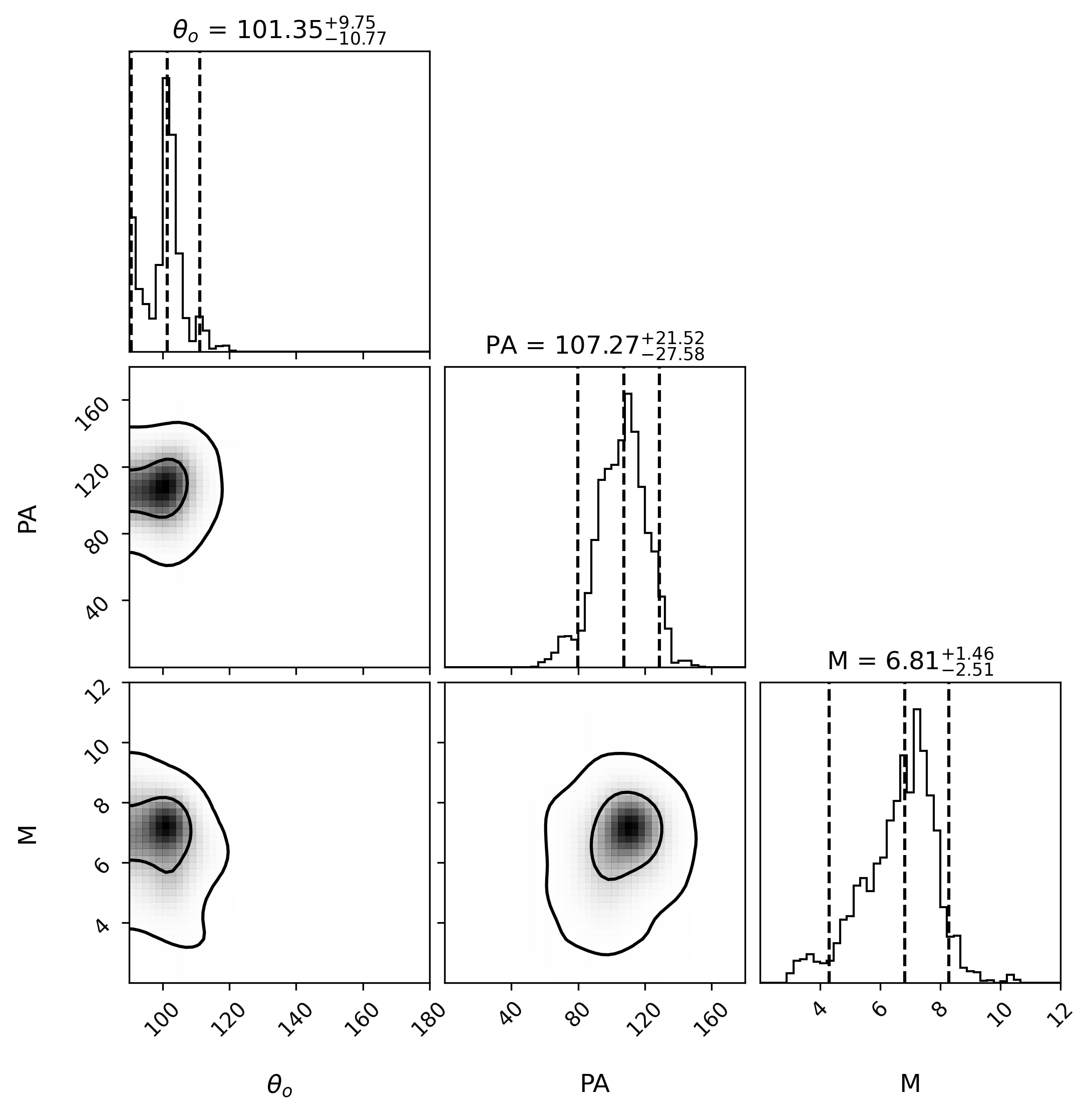}
	\includegraphics[width=0.38\textwidth]{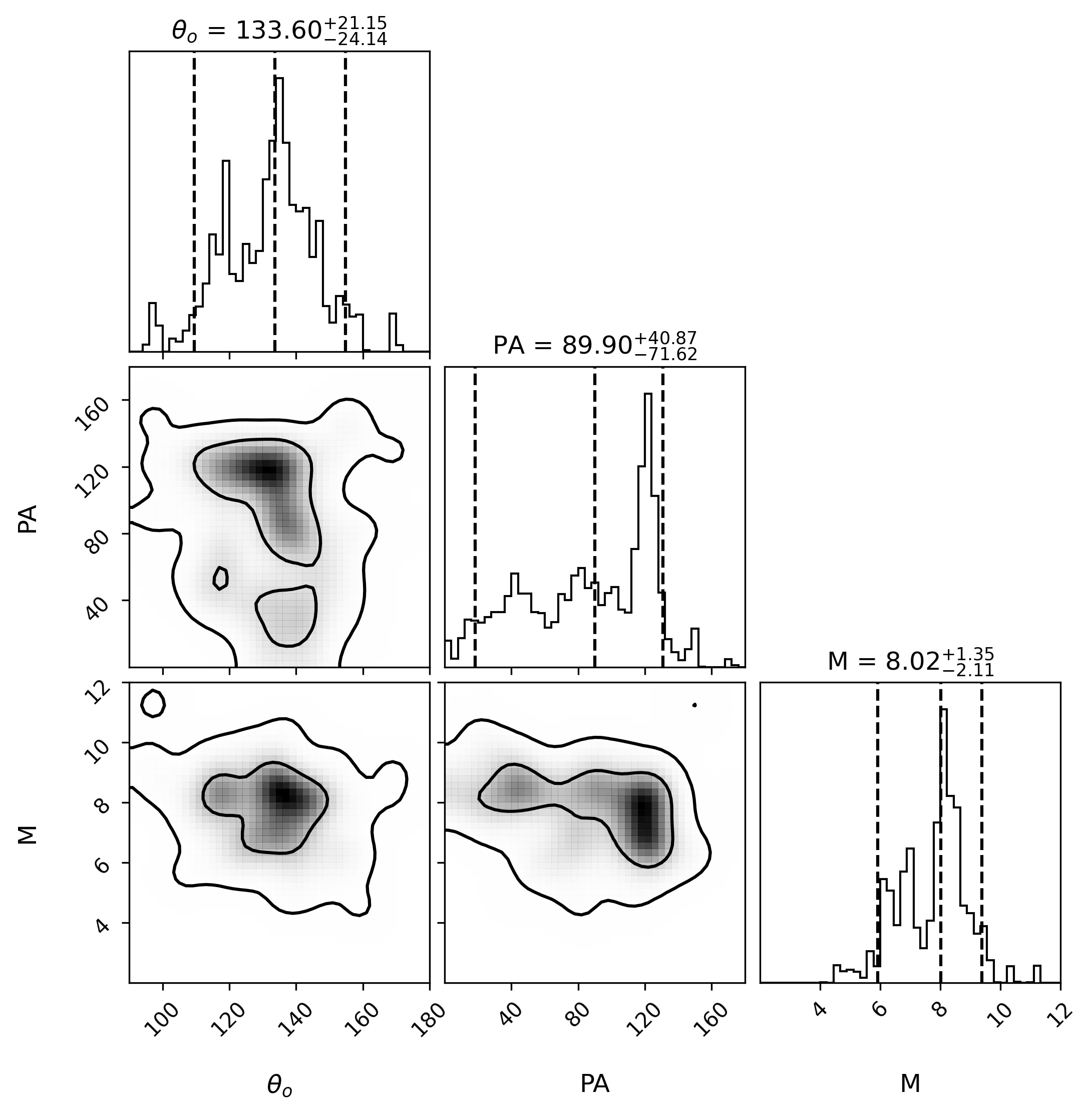}\\
	\includegraphics[width=0.38\textwidth]{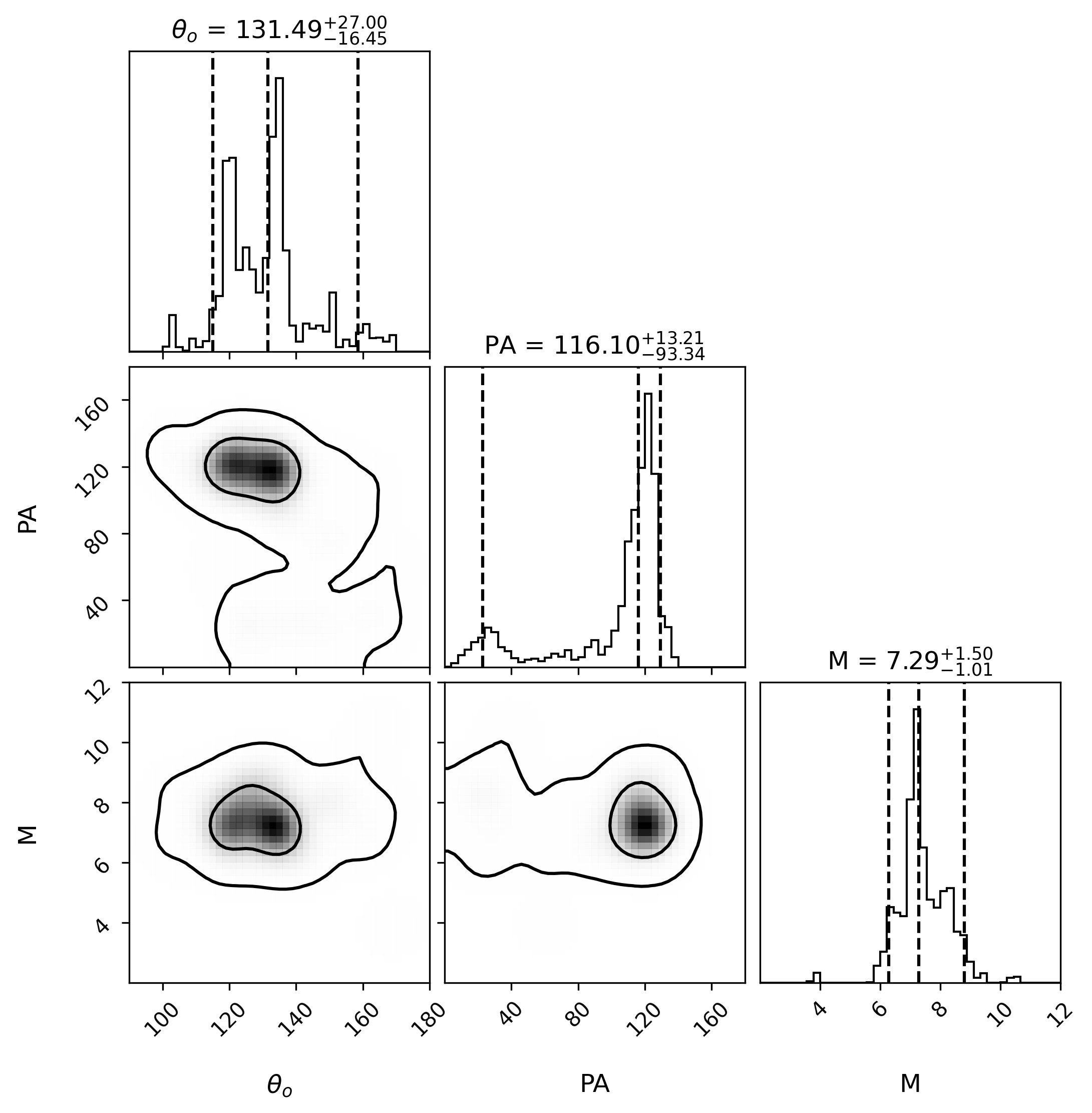}
	\includegraphics[width=0.38\textwidth]{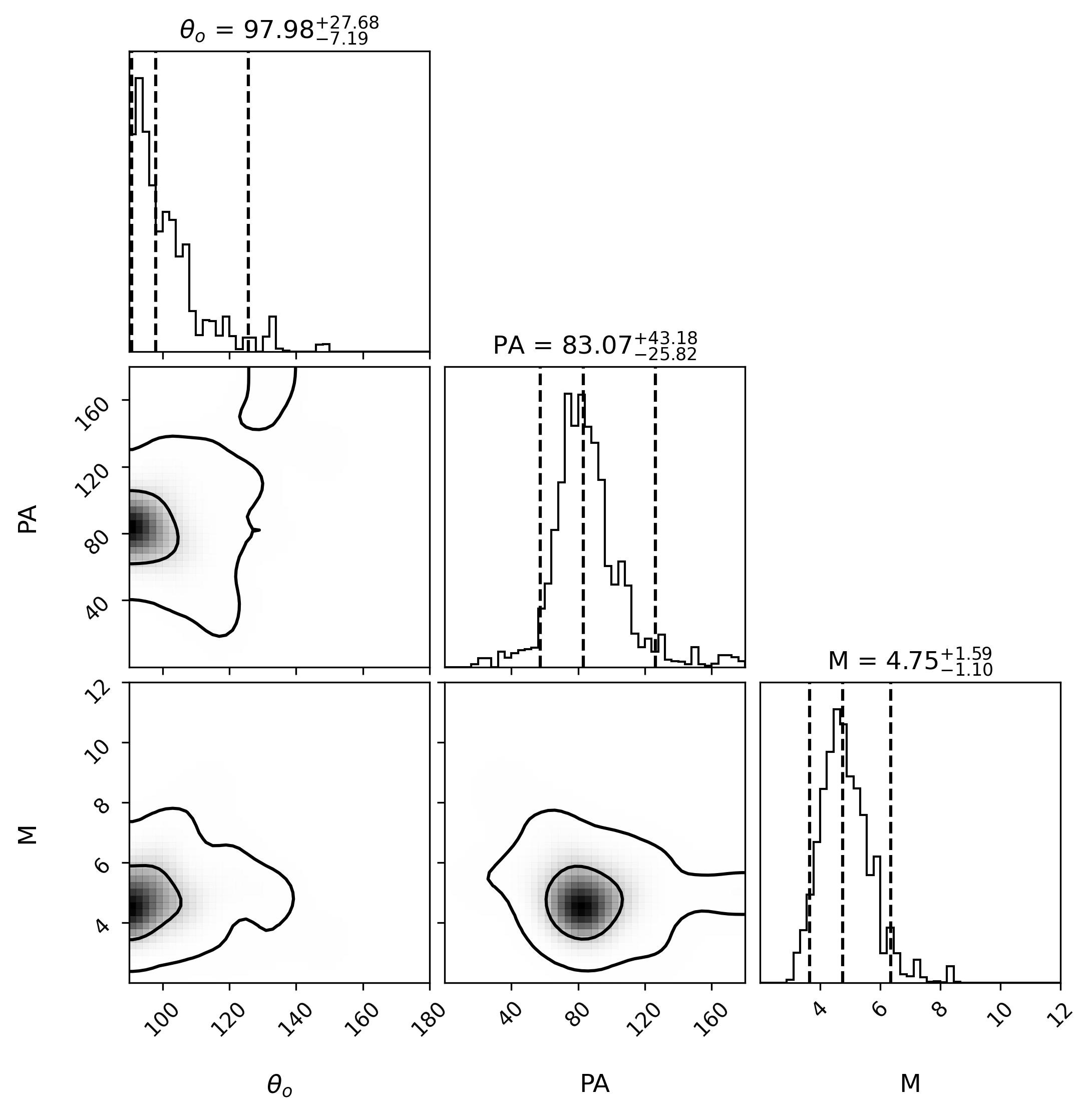}
	\caption{The constraint results from the joint analysis of the three events. The dashed lines in the diagonal plots represent the median values and the $90.0\%$ credible intervals of each parameter. The contours in the two-dimensional joint probability distribution plots correspond to the $39.4\%$ and $86.4\%$ credible regions. Top left: BS1. Top right: BS2. Middle right: BS3. Middle left: BS4. Bottom left: BS5. Bottom right: Schwarzschild BH.}
	\label{jointall}
\end{figure}

\subsection{Joint analysis of three events}

In this subsection, we present the MCMC results from the joint analysis of the three events. As noted earlier, the total dimensionality of the parameter space is nine for the joint analysis of three events. This prevents us from performing a $\chi^2$-grid evaluation, making it necessary to use MCMC sampling for the analysis. The posterior distributions, shown in the corner plots of Fig.~\ref{jointall}, illustrate the resulting constraints.

Owing to the substantially increased number of data points after combining the three events, the parameters are more tightly constrained compared to Fig.~\ref{bysBS2}. Our results show well-constrained mass distributions, with all marginal distributions for mass being single-peaked. For the position angle, half of the results are multi-peaked. Regarding the inclination angle, only BS1 and SBH exhibit single-peaked distributions, which are close to the equatorial plane, corresponding to a nearly equatorial observer inclination.

Our constraint on the mass of Sgr A* in the Schwarzschild case is $4.75^{+1.59}_{-1.10}\times10^6 M_\odot$. This value is consistent with other independent observational determinations, including $(4.297 \pm 0.0012)\times10^6 M_\odot$ inferred from stellar orbit measurements \cite{GRAVITY:2021xju} and $4.0^{+1.1}_{-0.6}\times10^6 M_\odot$ obtained by the Event Horizon Telescope from on the shadow size \cite{EventHorizonTelescope:2022exc}. However, similar to the results of combined data summarized in Tab. \ref{tab:combineddata}, the median ADM masses inferred for all boson star models are lower than in the Schwarzschild case. Among the five models, only the least compact BS1 model has its 90\% credible interval containing $4.3\times10^6 M_\odot$, the interval of BS3 just extends to this value, that of BS2 is slightly farther away, while the more compact BS4 and BS5 show substantial deviation from this value. This suggests that compact, spherically symmetric solitonic boson stars may not be suitable models for Sgr A*. In addition, as in the previous subsection, we again find no clear correlation between the inferred mass and the compactness of the boson star, indicating that the larger inferred masses are more likely caused by additional imaging features in boson star spacetimes that shift the observed centroid positions.

\subsection{ADM mass variation trend}
In the previous two sections, we constrained the combined data and the joint data using the grid method and the MCMC method, respectively, and obtained the corresponding results. In this section, we explore the relationship between the ADM mass of boson stars and their compactness based on these results.

\begin{figure}[!htbp]
	\centering
	\includegraphics[width=0.45\textwidth]{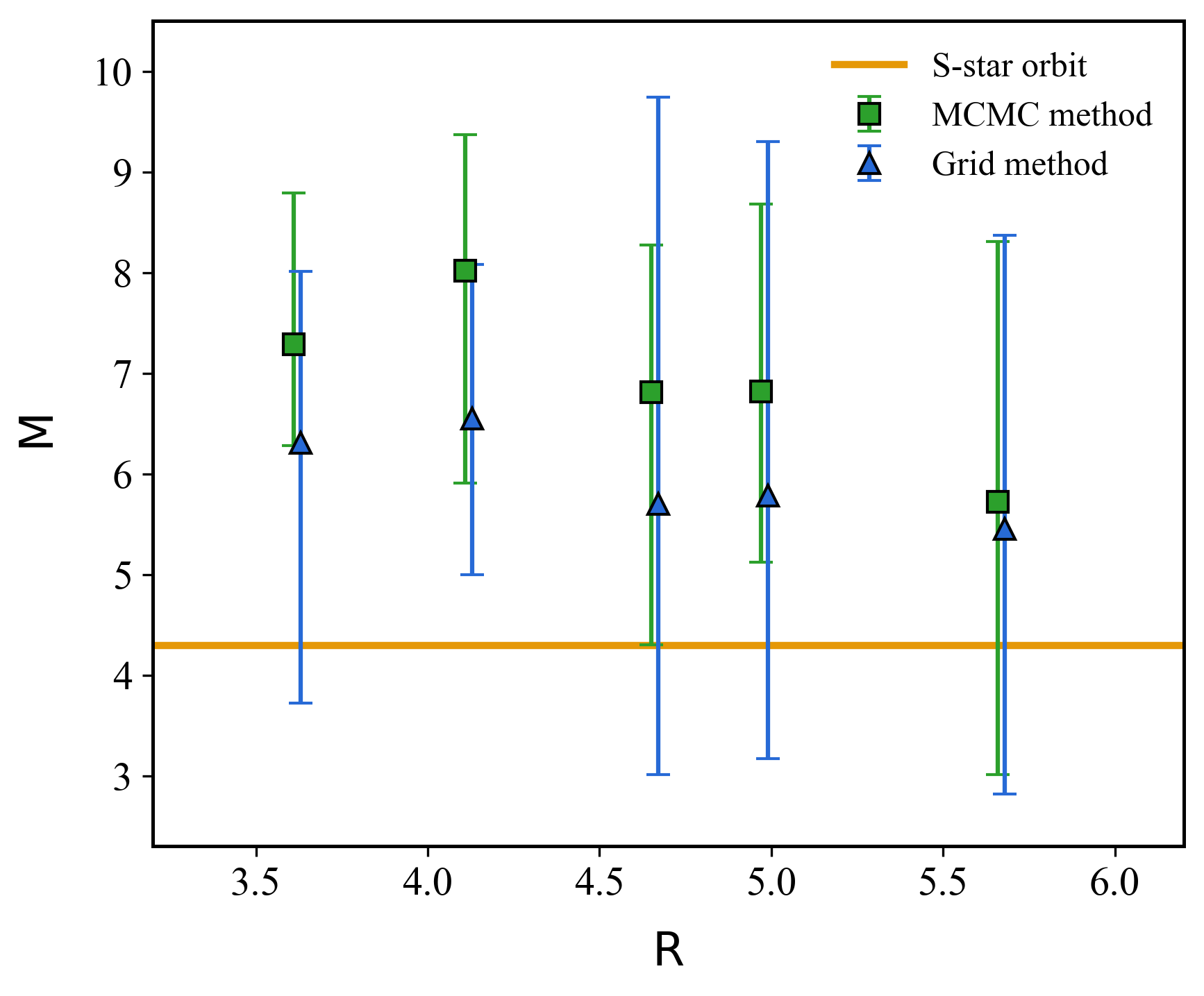}
	\caption{Boson star mass versus effective radius. The green square points represent the medians of the boson star masses constrained by the MCMC method, the blue triangles represent the medians of those constrained by the grid $\chi^2$ method, and the yellow region shows the constraint on the Galactic center mass by the S-star orbits, namely $(4.297 \pm 0.0012)\times10^6 M_\odot$ \cite{GRAVITY:2021xju}. The corresponding $90\%$ confidence intervals are marked with error bars in the figure.}
	\label{mr}
\end{figure}

Fig.~\ref{mr} presents the posterior distributions of the ADM mass derived from the two methods. The blue error bars correspond to the results obtained with the grid method, while the green error bars correspond to those obtained with the MCMC method. The medians are indicated by squares and triangles, respectively. To avoid confusion between the data from the two methods in the figure, a slight offset has been applied to the R‑coordinate of each point; this offset does not affect the overall trend. The yellow region shows the Galactic center mass value constrained by S-star orbits measurements. Because this constraint is much tighter than that derived from flare astrometry, it appears extremely narrow in the figure, effectively resembling a line.

From left to right, the five points correspond to five different configurations ordered by increasing compactness. First, it is worth noting that the confidence intervals of all inferred masses are larger than the S‑star value. However, for the more compact configurations on the left (i.e., BS5 and BS4), the yellow line only falls on the edge of the BS5 confidence interval obtained from the grid-$\chi^2$ method, while the MCMC result for BS5 and all results for BS4 fail to reach it. Second, for BS3 and BS2, the confidence intervals derived from the grid method all contain the accepted Sgr A$^*$ mass. Although the results obtained from the MCMC method do not enclose the yellow line, they are much closer compared to BS5 and BS4: the lower bound of BS3 is exactly $4.3\times10^6 M_\odot$, almost touching the yellow line.  Additionally, we note that the results for BS3 and BS2 are very close to each other, which we attribute to the strong similarity of their spacetime metrics. As can be seen in Fig.~\ref{so}, although the matter field of BS3 approaches zero more rapidly, the spacetimes of BS2 and BS3 are nevertheless very similar. Finally, we focus on the most diffuse configuration, BS1, for which both the blue and green error bars contain the accepted value. Among all configurations, the median of BS1 is the closest to the established value. From this, we identify a trend: more diffuse boson star configurations yield constraints that more closely resemble those of a black hole, which is contrary to our intuition.

\subsection{From the image perspective}

\begin{figure}[!htbp]
	\centering
	\includegraphics[width=0.95\textwidth]{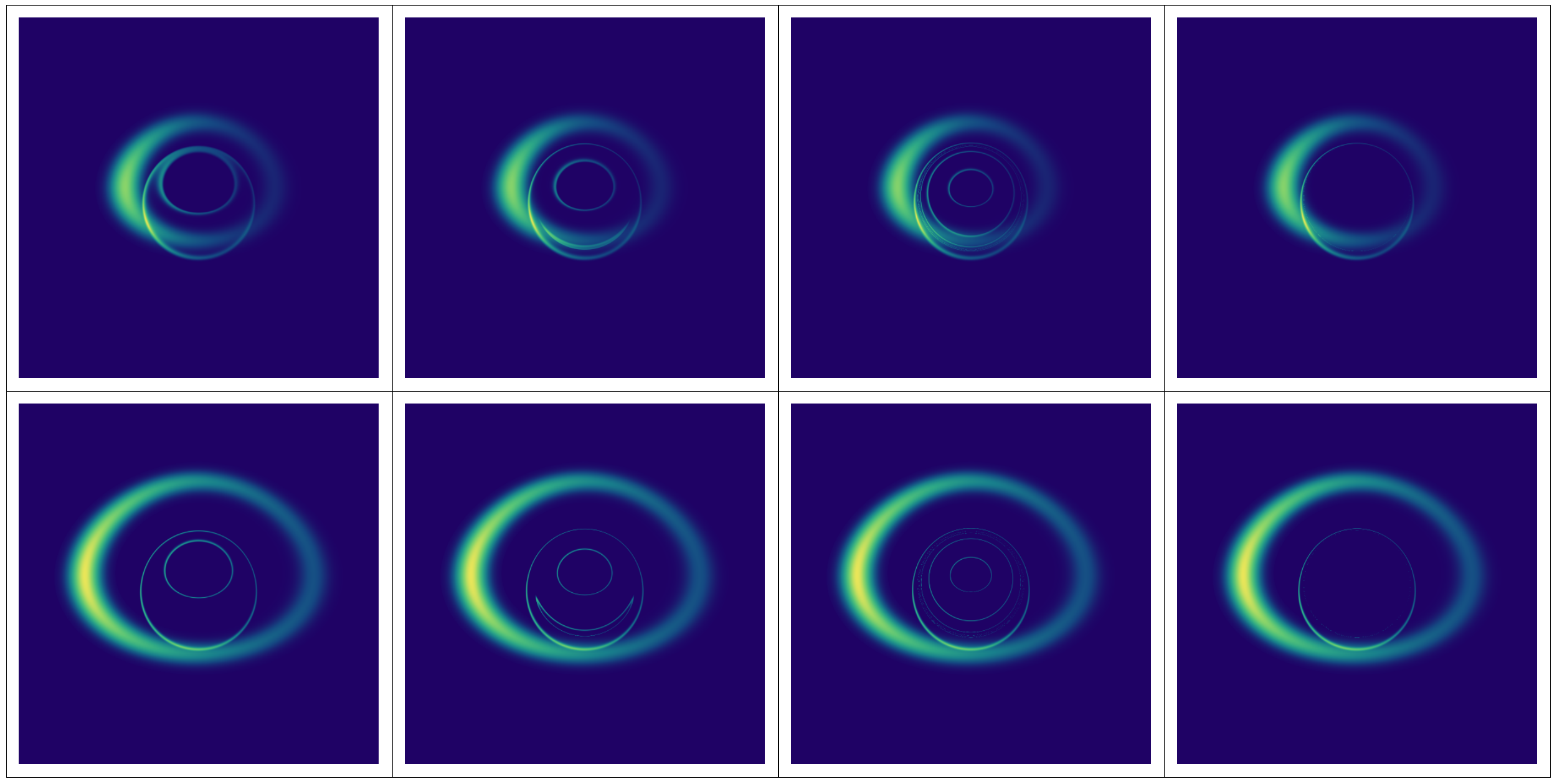}
	\caption{From left to right are the time-averaged imaging examples of BS1, BS3, BS5, and SBH. We choose the inclination $\theta_o = 130^\circ$. The top row corresponds to $r_\text{hs} = 6M$, and the bottom row corresponds to $r_\text{hs} = 10M$.}
	\label{imgs}
\end{figure}

To understand the origin of the differences between the boson star results and the Schwarzschild case, we show the time-averaged images for the various models, as illustrated in Fig.~\ref{imgs}. We find that more compact boson star models produce a larger number of additional images relative to the Schwarzschild case. This is natural since a highly compact boson star (e.g., BS5) bends the light rays passing through the object more strongly, allowing multiple higher-order images to form, whereas a less compact boson star (e.g., BS1) produces only a single image generated by rays traversing the star. The lensed images lie closer to the center of the image, so their presence reduces the radius of the brightness centroid. Therefore, although BS1 exhibits the largest deviation from Schwarzschild in its metric components, its imaging appearance is the most similar to Schwarzschild. As a result, the inferred parameters such as mass and inclination turn out to be the closest to those of the Schwarzschild case.

According to our fits, the Schwarzschild spacetime consistently yields the largest hotspot radius and the smallest ADM mass across all single event datasets. These two parameters are partly degenerate: increasing the hotspot orbital radius enlarges the centroid radius of the brightness distribution, while increasing the ADM mass uniformly scales up the entire image, which also increases the centroid radius. However, the images themselves reveal how these two effects differ. Comparing scenarios with different values of $r_\text{hs}$, we find that varying $r_\text{hs}$ directly alters the size of the primary image, whereas the locations of lensed images vary only slightly. In other words, modifying the hotspot orbital radius shifts the relative separation between the primary and higher-order images, while changing the mass does not.

\section{Summary and discussion}\label{sec6}

In this study, we tested whether Sagittarius A* could be a solitonic boson star by using GRAVITY NIR flare astrometry data. We chose five different boundary conditions and potential parameters, and numerically solved the equations of motion, thereby obtaining the metrics of five spherically symmetric solitonic boson stars with different compactness, i.e., with different effective radii. By interpreting the flares as hotspots undergoing circular geodesic motion around Sgr A*, we used GRRT to generate dynamical images for different models at various hotspot orbital radii and inclinations. From these dynamical images, we computed the time evolution of the intensity centroid positions and compared them with the observational data. For a more direct comparison, we also carried out the same procedure for the Schwarzschild black hole case. The fitting procedure involves both global parameters shared among different flare events, including the inclination, central object mass, and position angle, and event-specific orbital parameters describing the hotspot motions. Using a grid-based $\chi^2$-evaluation approach, we analyzed the combined data for all astrometric flare events released by GRAVITY. For the global parameters, we also performed a joint Bayesian inference of multiple events using MCMC methods.

For the $\chi^2$ results based on the combined data, nearly all boson star models tend to yield larger masses and smaller hotspot orbital radii than the Schwarzschild black hole model. This systematic preference for larger masses also appears in the MCMC results from the joint analysis of the three events. For the MCMC results, only the most diffuse configuration has a 90\% credible interval that includes the known mass of Sagittarius A* from other observations. Overall, the GRAVITY flare astrometry tends to favor an interpretation of Sgr A$^*$ as a spherically symmetric black hole rather than a solitonic boson star, although a less compact boson star remains a viable possibility.


In addition, the inferred mass does not exhibit a clear monotonic dependence on compactness, which could be more naturally attributed to the transparency of boson stars rather than to their metric deviations. Owing to the absence of an event horizon, boson stars are transparent to light, allowing photons to traverse their interiors and generate additional higher-order images. These interior images typically appear closer to the center of the brightness distribution and pull the centroid inward, thereby significantly modifying the predicted centroid motion and strongly affecting the fitting results. More compact boson star models tend to produce multiple such interior images, whereas less compact models usually generate only one. As a consequence, although the less compact boson stars show more deviation from the Schwarzschild metric, their imaging signatures are comparatively closer to those of a black hole, leading to inferred parameters that more closely resemble the Schwarzschild case.

Future works may concentrate on several directions. Firstly, it would be interesting to explore direct constraints on the boson-star parameters, namely $\alpha$ and $\psi_0$. This requires computing GRRT images for multiple parameter choices, which substantially increases the dimensionality of the inference and therefore the computational cost. Secondly, it is natural to extend our study to axisymmetric cases, which could be more realistic. Given that the spherically symmetric boson star considered here yields a relatively large mass discrepancy, it would be worthwhile to investigate rotating boson stars and assess how spin affects the mass constraints. Thirdly, GRAVITY polarization data are more abundant than astrometric measurements, so incorporating the observed Q–U loops to constrain the model could potentially provide stronger and more robust results.

\section*{Acknowledgments}
We thank Jiewei Huang, Zelin Zhang and Qing-Hua Zhu for insightful discussions. The work is partly supported by NSFC Grant No. 12575048, 12275004, 12205013, 12547123, 12588101 and 12547127. M. Guo is also supported by Open Fund of Key Laboratory of Multiscale Spin Physics (Ministry of Education), Beijing Normal University and he is also supported by the BNU Tang Scholar.

\bibliographystyle{utphys}
\bibliography{gravityflare}

\end{document}